\documentclass[fleqn,10pt]{wlscirep}
\usepackage[utf8]{inputenc}
\usepackage[T1]{fontenc}
\usepackage{xr}
\usepackage{wrapfig}
\title{The Drivers of Global News Spreading Patterns}

\author[1]{Shayan Alipour}
\author[2]{Niccolò Di Marco}
\author[1]{Michele Avalle}
\author[1]{Gabriele Etta}
\author[1]{Matteo Cinelli}
\author[1,*]{Walter Quattrociocchi}
\affil[1]{Sapienza University of Rome, Italy}
\affil[2]{University of Florence,  Italy}

\affil[*]{walter.quattrociocchi@uniroma1.it}

\begin{abstract}

The web radically changed the dissemination of information and the global spread of news. In this study, we aim to reconstruct the connectivity patterns within nations shaping news propagation globally in 2022. 
We do this by analyzing a dataset of unprecedented size, containing 140 million news articles from 183 countries and related to 37,802 domains in the GDELT database.
Unlike previous research, we focus on the sequential mention of events across various countries, thus incorporating a temporal dimension into the analysis of news dissemination networks.
Our results show a significant imbalance in online news spreading. 
We identify news superspreaders forming a tightly interconnected rich club, exerting significant influence on the global news agenda.
To further investigate the mechanisms underlying news dissemination and the shaping of global public opinion, we model countries' interactions using a gravity model, incorporating economic, geographical, and cultural factors.
Consistent with previous studies, we find that countries' GDP is one of the main drivers to shape the worldwide news agenda.

\end{abstract}

\begin{document}

\flushbottom
\maketitle

\thispagestyle{empty}

\section*{Introduction}

The rapid evolution of the web has profoundly transformed the global dissemination of news. Nowadays, news flows incessantly, with media outlets promptly reporting unfolding events, and anyone can access this real-time information ecosystem. Furthermore, the Internet has intensified competition among news providers \cite{Ahmad2010, Gentzkow2008} as they strive to capture users' attention, contending with both traditional media and online news sources \cite{chen2019competition}.
Advancements in data science have unlocked new possibilities for analyzing vast amounts of news content \cite{leetaru2012data, sittar2020dataset} with extensive research efforts devoted to studying the impact of social media platforms on news dissemination \cite{bodaghi2022theater, abdullah2011epidemic, schmidt2017anatomy}. A particular emphasis has been posed on their influence on misinformation \cite{del2016spreading, vosoughi2018spread} and polarization \cite{cinelli2021echo, spohr2017fake, stroud2010polarization}.
In this dynamic environment, news spreads rapidly across the web, reaching numerous outlets in different countries and potentially shaping the news agendas of other nations \cite{wanta2004agenda, kiousis2008international}. Nevertheless, the reach and influence of the global news network vary significantly from one country to another. Factors such as culture, government censorship, language barriers, and digital division can impact news exchange between nations \cite{peter2003agenda}.
Therefore, gaining a comprehensive understanding of how news circulates on the Internet yields valuable insights into the economic and cultural influences among countries. It allows for characterizing news spreading routes, providing insights into the interconnected relationships and interdependence between different nations.
Previous studies \cite{wallerstein1974wst, Chang1998} have shed light on the key factors influencing news dissemination, including economic strength, geographic location, and cultural elements. These factors are pivotal in shaping the pathways through which news flows across countries. Additionally, studies \cite{kim1996network, wu2003homogeneity, gdelt_network_2022, gravitynews320, Fracasso2014} have revealed hierarchical structures and clustering patterns that are driven by economic growth, language, and political freedoms. Such findings emphasize the complex dynamics in the global news network, where certain countries or regions may have stronger connections due to shared economic interests, linguistic affinities, or similar political environments.
By analyzing the patterns of news circulation, we can map the intricacies of global information exchange, gaining insights into how countries interact and influence each other through news dissemination. These insights contribute to a deeper understanding of the multifaceted relationships in the modern interconnected world.
In this study, we aim to construct a comprehensive global network of news flows by leveraging the Global Database of Events, Language, and Tone (GDELT) for 2022 \cite{leetaru2013gdelt, leetaru2015mining}.
Our analysis focuses on the order in which news spreads within media outlets, moving beyond a simple examination of the frequency of countries' names in news articles. 
By studying the sequential mentions of events across countries, we aim to map the intricate interactions and reciprocal influences among nations in disseminating news.
To achieve this, our methodology involves identifying the initial source of each news event and tracing subsequent mentions of the event by other countries. By doing so, we can unravel the paths through which news travels across the global media landscape. This approach allows to gain insights into how news events unfold, propagate, and are shared among countries, shedding light on the dynamics of news dissemination and the interconnectedness of nations in shaping the global information landscape.
We leverage a massive dataset of 140 million news articles from 37,802 domains across 183 countries sourced from GDELT. 
We find a significant skewness and inequality in the resulting graph, representing the interconnectedness of news dissemination. Specifically, we observe a small set of dominant countries that form a well-connected network, indicating their influential role in disseminating news globally and shaping the global public agenda.
In pursuit of a deeper comprehension of the intricate dynamics underlying the patterns of news diffusion across countries, we use a gravity model. This methodological approach allows us to unravel the fundamental drivers of the observed skewed distribution. Our analysis reveals two primary determinants that significantly influence this phenomenon: the gross domestic product (GDP) and the geographic proximity of the countries involved. Notably, nations endowed with larger GDPs and closer geographical proximity demonstrate heightened interconnections and wield substantial influence within the global news network. Furthermore, we acknowledge the salient role common languages play, which actively shapes the flow of news across nations.

\section*{Results}

\subsection*{News and Countries}
\begin{figure}
    \centering
    \includegraphics[scale = 0.6]{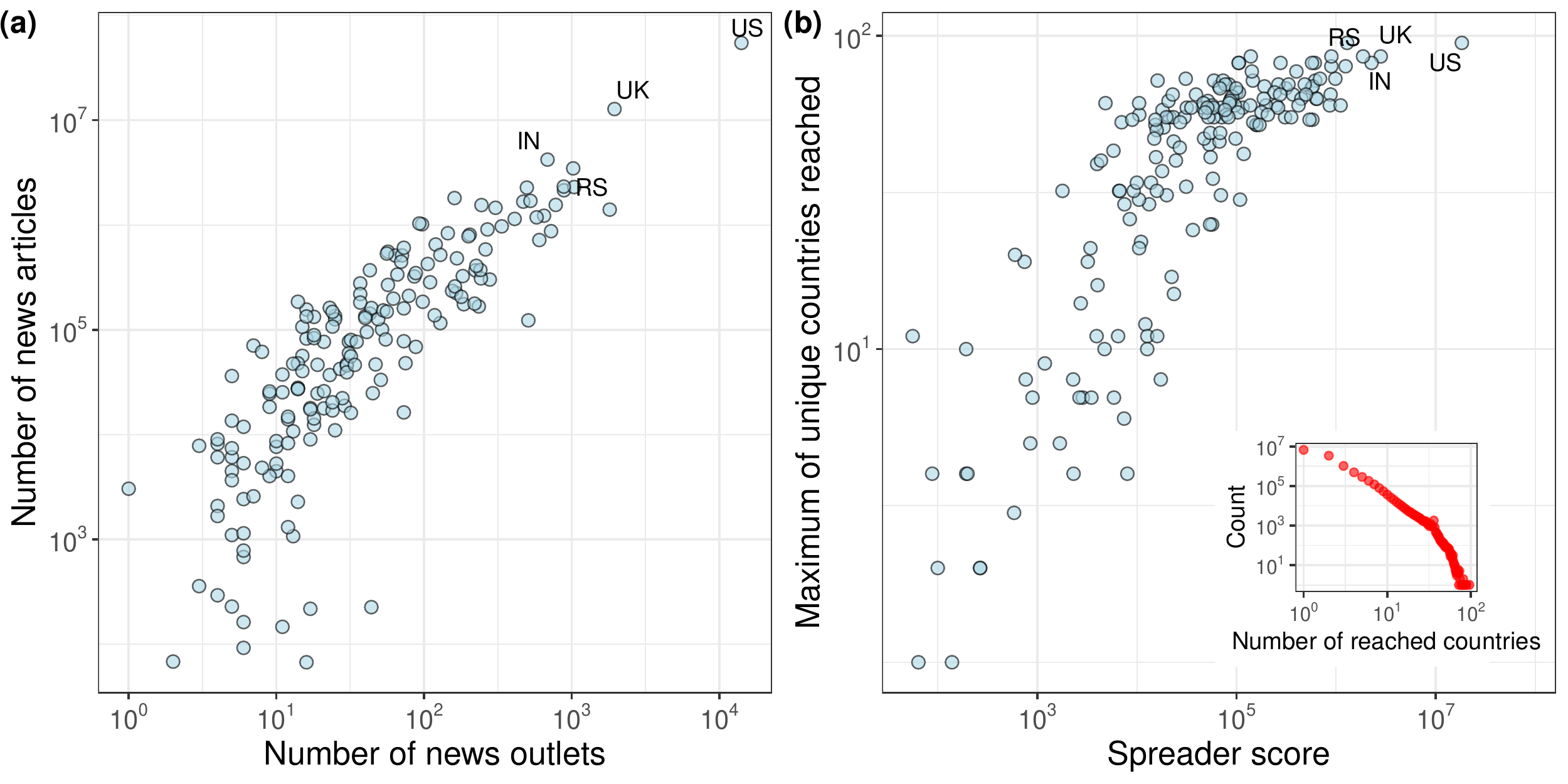}
    \caption{Left panel $(a)$ describes the number of associated news outlets and published news articles for each country. Right panel $(b)$ shows how many times a country served as the initial news spreader of the event and the maximum virality of events spread by each country. The inset in $(b)$ reports the virality distribution (i.e. the number of unique countries an event gets mentioned in) for all events regardless of the initial spreader country.}
    \label{fig:data_info}
\end{figure}

To assess the extent of news outlet activity across different countries, we rely on the Source-Country dataset, as detailed in the Methods section, which allows us to identify and analyze news outlets alongside their respective news articles. To visualize the relationship between the number of news outlets and the corresponding quantity of news articles published by each country, we present Figure \ref{fig:data_info}(a). This graphical representation depicts the interplay between these two variables and shows that countries with more news outlets tend to publish more news articles, following approximately a power law.

We define the initial spreader(s) as the country or countries that first mention the event. Since the GDELT dataset has a time granularity of 15 minutes, multiple countries can be identified as the initial spreaders for a single event. This is because two or more news outlets may have posted an article within the same 15-minute window. 

For each event, we consider the number of countries in which that event was mentioned. We call the distribution of these measures the \textit{virality} distribution. Moreover, we define the \textit{spreader score} for a country $i$ as the number of times $i$ belongs to the set of initial spreaders of an event. Figure \ref{fig:data_info}(b) depicts the spreader score of a specific country plotted against the maximum number of countries reached by events published by that country.

Figure \ref{fig:data_info}$(b)$ indicates that the United States (US) emerged as the initial top spreader in 2022, significantly surpassing other countries. The United Kingdom (UK) and India (IN) followed closely behind.
Moreover, we deduce that countries with higher spreader scores are more likely to have their events go viral at least once. The inset plot in Figure \ref{fig:data_info} illustrates the overall distribution of virality, suggesting that most events tend to circulate within their countries.

\subsection*{Structural Analysis on News Spreading}
\begin{figure}[!ht]
    \centering
    \includegraphics[scale = 0.85]{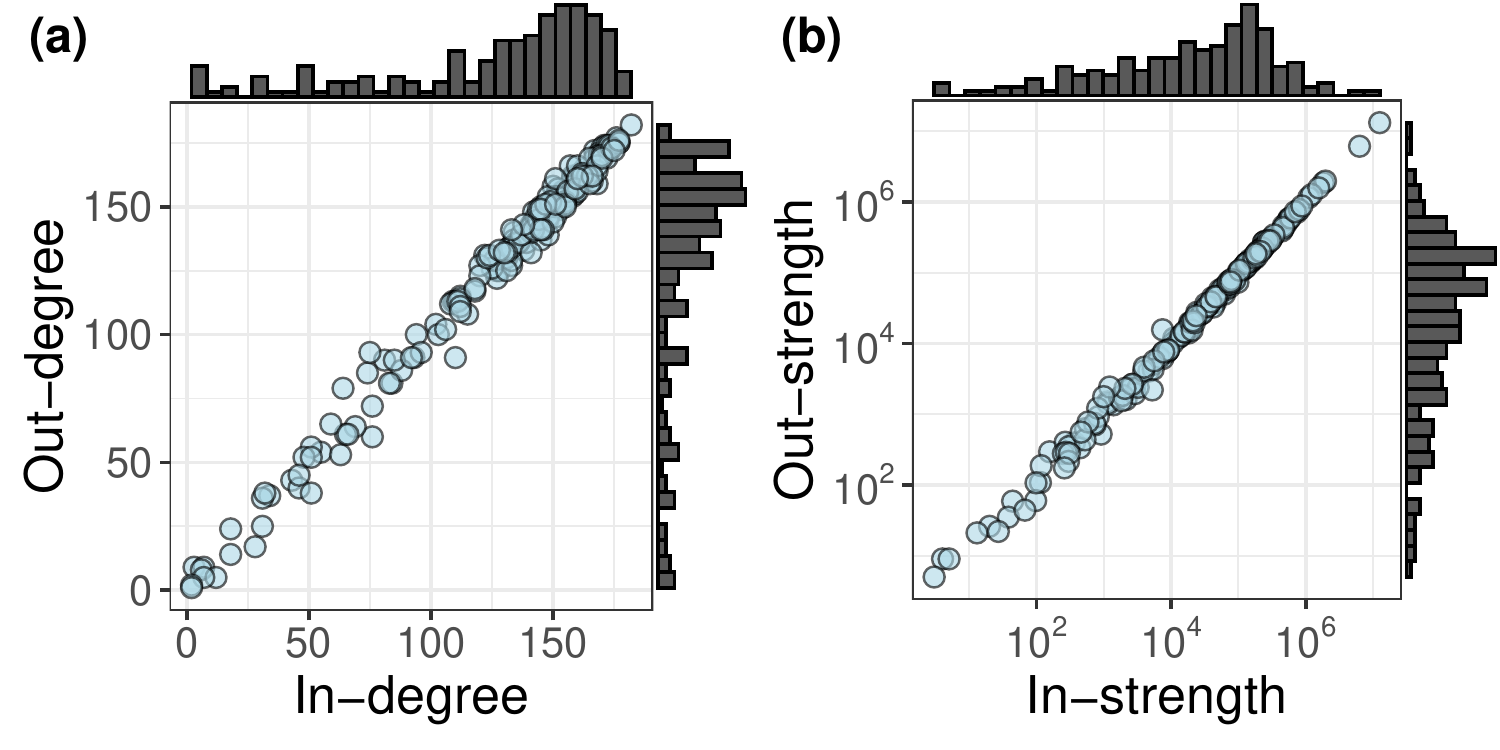}
    \caption{$(a)$ relation between in and out-degree and $(b)$ relation between in and out-strength. In both cases, they are highly correlated.}
    \label{fig:deg_distribution}
\end{figure}

In this section, we aim to quantify the heterogeneity in the news-spreading process. We construct a directed weighted network $G = (V, E)$ using the GDELT data to achieve this. The set of countries, denoted as $V$, represents the nodes in the network. A directed edge $(i, j) \in E$ is created if country $j$ mentions an event successively after country $i$. The weight associated with each edge, denoted as $w_{ij}$, represents the number of times this sequential mentioning occurs. Please refer to the Methods section for more detailed information on this construction.
The resulting network consists of 183 countries and 23,033 edges. The network's density, which indicates the proportion of existing connections compared to all possible connections, is 0.69. This value suggests that relatively few pairs of countries lack an exchange of information.
Figure \ref{fig:deg_distribution} illustrates the relationship between in-degree and out-degree, as well as in-strength and out-strength within the network.

The degree distributions, both in-degree, and out-degree, reveal that most countries in the network have many connections. This indicates that countries tend to be mentioned by many other countries (in-degree) or mention many other countries (out-degree) at least once in the news-spreading process. It suggests that countries are actively involved in disseminating news and have significant interactions with other countries.

The in-strength and out-strength distributions exhibit a similar pattern but on a logarithmic scale. This indicates that while most countries have a relatively high level of involvement in receiving or transmitting information, a few stand out with exceptionally high levels of strength in news dissemination. These countries play a prominent role in spreading the news, indicating a concentration of influence in the network.

An interesting observation is a high correlation between the in-degree and out-degree ($\rho = 0.99, p < 2.2 \times 10^{-16}$). This can be justified by the high reciprocity (i.e. proportion of links in both directions) exhibited by the network, with a value of 0.91.
Furthermore, when comparing the weights of mutual edges, they tend to have similar values. This implies that the extent of news diffusion between countries is well-balanced, as indicated by the similarity in the weights of reciprocal edges. Please refer to Figure~\ref{fig:mutual_edges} in the supplementary information for additional details and visualizations.

To examine the organizational principles of the network, we analyze its topology concerning the associated weights. In line with previous studies \cite{Barrat2004}, we compare the unweighted and weighted clustering coefficients to gain insights into the correlation between weights and network topology. Specifically, we define $C_w (s)$ as the weighted directed clustering coefficient \cite{Clemente2018} and $C (s)$ as its unweighted counterpart, both averaged over nodes with a strength of $s$ (see Methods for details about the coefficients).

\begin{figure}[!ht]
    \centering
    \includegraphics[scale = 0.75]{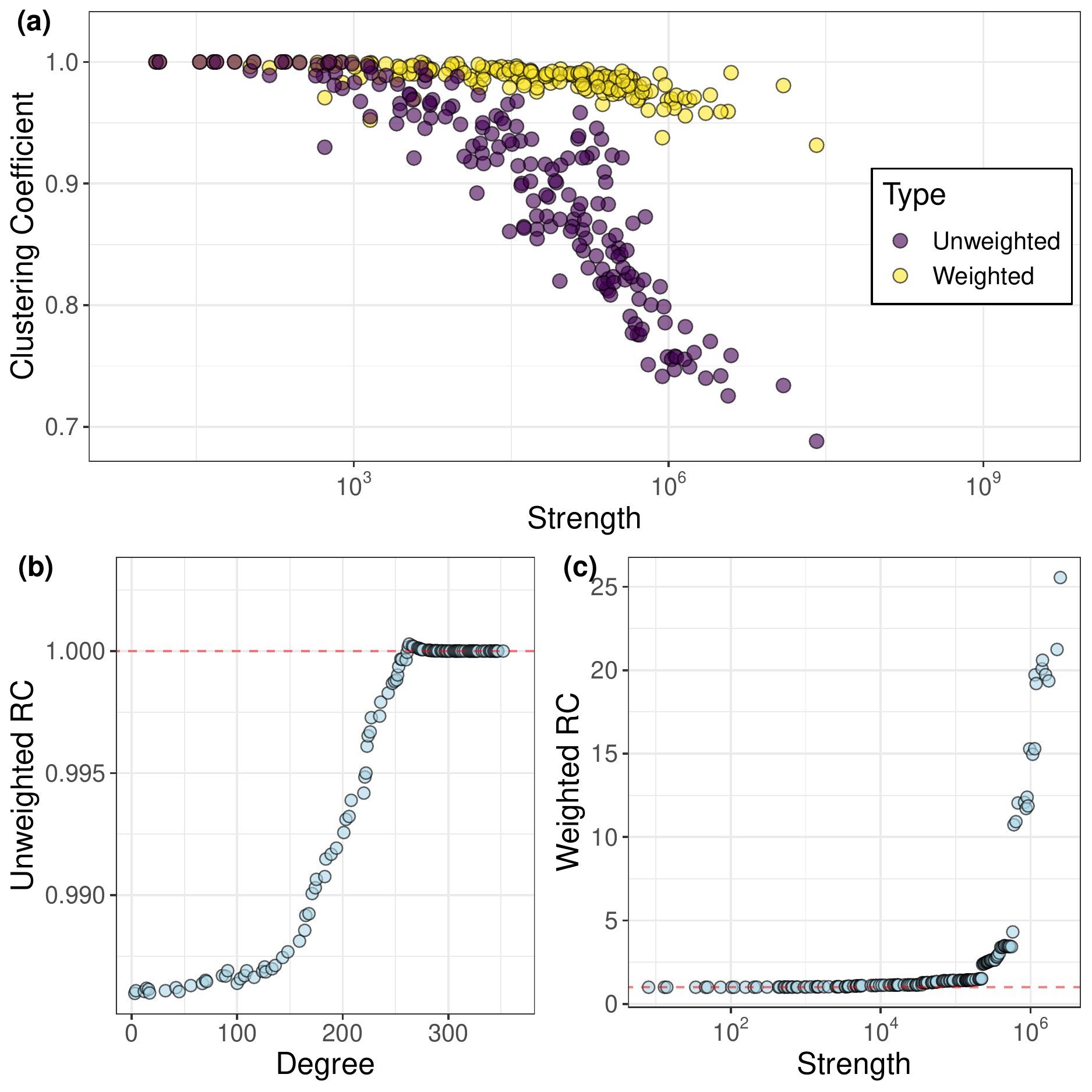}
    \caption{$(a)$: Comparison between unweighted and weighted clustering coefficient. The weighted one takes higher values for higher strength. 
    $(b)$: Normalized rich-club (RC) coefficient for each degree value. $(c)$: Normalized weighted rich-club coefficient for each strength value.}
    \label{fig:rc}
\end{figure}

Figure~\ref{fig:rc}$(a)$ shows the results of the comparison of $C_w (s)$ and $C (s)$ as the strength increases.

The weighted directed clustering coefficient $C_w (s)$ is approximately equal to the unweighted clustering coefficient $C(s)$ for low-strength nodes. However, $C_w (s)$ takes on higher values for high-strength nodes than $C(s)$. This finding suggests that while higher-strength nodes tend to form fewer triangles in terms of network topology, the triangles they do form are more likely to consist of edges with higher weights. In other words, there is a core group of countries where information flows preferentially, contributing to higher weighted clustering coefficients. This indicates the presence of strong interconnectedness and concentrated information exchange within this core group of countries. This is also suggested by Figure~\ref{fig:heatmap} in supplementary information.

The observed scenario in terms of clustering coefficient and the observed in and out-strength distributions form the basis for detecting a rich-club phenomenon \cite{xu2010rich}. A rich club refers to a group of prominent and tightly interconnected nodes that exert control over the flow of information in the network \cite{zhou2004rich, Barrat2004, Opsahl_rich_club, cinelli2019generalized}.
Figure \ref{fig:rc} compares the normalized unweighted and weighted rich-club coefficients (refer to the Methods section for further details). The normalized rich-club coefficient measures the extent to which high-degree nodes tend to be more interconnected with each other compared to what would be expected by chance taking into account an appropriate null model.
Consistent with the previous findings, we observe that only a strong weighted rich-club ordering (with the coefficient significantly higher than 1 for high-degree nodes) is evident, confirming our earlier hypothesis. Specifically, the eight largest countries (according to their strength) within the rich-club are the United States (US), the United Kingdom (UK), Canada (CA), India (IN), Russia (RS), Germany (GM), France (FR), and Ukraine (UP). These countries play a prominent role in the weighted network, forming a core where information flow is significantly concentrated and interconnected. Contrary to expectations, it is noteworthy that Ukraine belongs to the rich-club. This is likely due to the extensive coverage of events concerning the war between Russia and Ukraine in 2022, contributing to their prominent position in the network.

\begin{figure}[!ht]
    \centering
    \includegraphics[width = \linewidth]{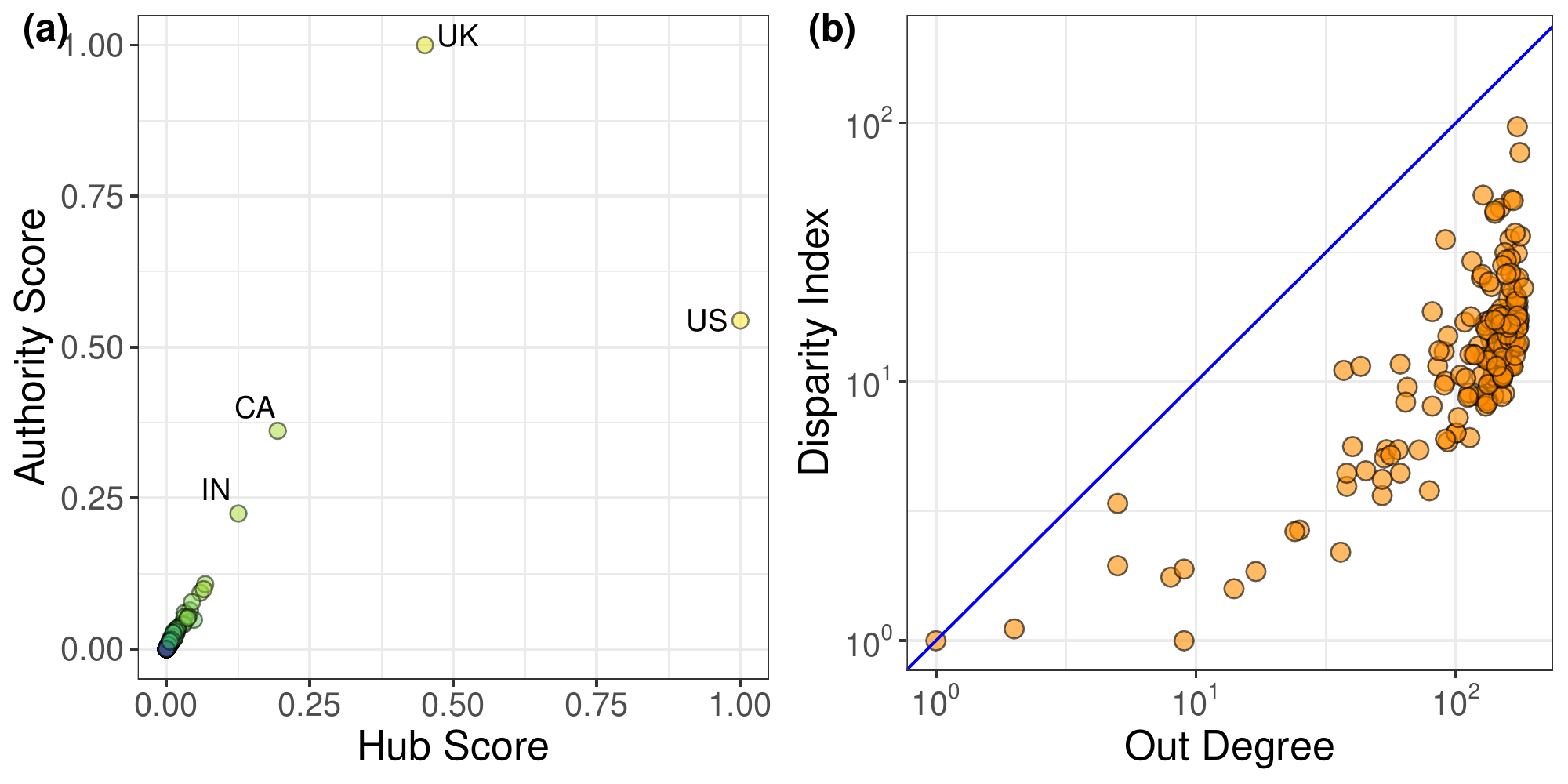}
    \caption{(a) Comparison between (weighted) Hub and Authority scores. We observe that US dominates as a Hub of the Network, while UK is the biggest Authority.
    (b) Disparity index $\gamma_i (k)$ versus the out degree of network' nodes. We observe an increasing value of $\gamma_i (k)$ when the out Degree is greater.}
    \label{fig:hits}
\end{figure}

All of the previous analyses indicate that only a small set of countries is actively involved in news diffusion, while the participation of other countries is relatively minimal. However, it is also interesting to examine the individual role of each country in the network. To accomplish this, we employ the HITS algorithm \cite{Kleinberg1999} and compare the Authority and Hub scores of each node in Figure \ref{fig:hits}, utilizing the weights of the edges to represent the intensity of interaction. The color of the nodes corresponds to their strength.

The plot confirms that the United States (US), the United Kingdom (UK), Canada (CA), and India (IN) have a significant influence on news diffusion. Specifically, the United States emerges as the primary hub in the network, indicating its central role in disseminating news. Conversely, the United Kingdom assumes the primary authority position, suggesting its influence in shaping the information landscape.
Canada and India play a mixed role, exhibiting authority and hub characteristics to some extent. However, it is important to note that the global domination of news diffusion conversation is primarily attributed to the United Kingdom and the United States. These countries play central and influential roles in disseminating news on a global scale.

\subsection*{Heterogeneity of diffusion and gravity model}
In the previous section, we demonstrated that only a small group of countries actively participate in the dissemination of information, indicating a high degree of heterogeneity in how news is spread. To further quantify this aspect, we proceed by computing transition probabilities associated with each edge., i.e.

\begin{equation}\label{eq:transition_prob}
    p_{ij} = \frac{w_{ij}}{\sum_{k = 1}^{deg(i)} w_{ik}}.
\end{equation}

The transition probability $p_{ij}$ represents the proportion or percentage of information produced by country $i$ that flows to the country $j$. We are primarily interested in understanding if the observed behavior could emerge randomly.
In Figure~\ref{fig:hits}$(b)$, we calculate the disparity index $\gamma_i (k)$ \cite{Serrano2009} (see Methods) and display it. The blue line indicates the maximum value that $\gamma_i (k)$ can assume. The values observed in Figure~\ref{fig:hits}$(b)$ indicate that news diffusion heterogeneity increases with the node's out-degree. This suggests that news diffusion does not occur randomly or uniformly across the network. If random, we would expect a more consistent behavior, resulting in $\gamma_i (k) \approx 1$ for any given $k$.
The increasing values of $\gamma_i (k)$ concerning the out-degree imply that certain countries play a more influential role in news diffusion than others. This further supports the notion that news spreading is characterized by heterogeneity, where some countries have a more significant impact and influence in disseminating information than others.

We propose using a gravity model (GM) applied explicitly to news dissemination to investigate the factors influencing the observed behaviour in news diffusion. This approach has been employed in some prior studies, albeit with slight variations in network considerations \cite{Fracasso2014}.
For our analysis, we utilize the CEPII gravity dataset \cite{conte2021cepii}, which provides comprehensive economic data on international trade and various aspects of the global economy. However, since this dataset is updated only until 2020, we merge it with GDP data from 2022 obtained from the International Monetary Fund (IMF)\cite{IMF}.
Doing so, we aim to incorporate relevant economic factors into the gravity model for news diffusion. This approach will allow us to examine how economic variables, such as GDP, influence news flow between countries.

Our considered model is:

\begin{equation}
    logit(p_{ij}) = \beta_0 + \beta_1 log(GDP_i) + \beta_2 log(GDP_j) + \beta_3 log(d_{ij}) + \beta_4 C_{ij} + \beta_5 L_{ij} + \beta_6 S_{ij},
\end{equation}

where:

\begin{enumerate}
    \item $GDP_i$ is the GDP of country $i$ measured in billions of dollars;
    \item $d_{ij}$ is the distance between the most populated cities of the two countries $i$ and $j$, measured in $km$;
    \item $C_{ij}$ is a dummy variable equal to $1$ when country $i$ and $j$ share a common border;
    \item $L_{ij}$ is a dummy variable equal to $1$ when country $i$ and $j$ share a common language;
    \item $S_{ij}$ is a dummy variable equal to $1$ when country $i$ and $j$ share a common language spoken by at least $9\%$ of the population;
\end{enumerate}

Note that our dependent variables are the logit of $p_{ij}$, namely

\begin{equation}
    logit(p_{ij}) = \log \left( \frac{p_{ij}}{1-p_{ij}}\right),
\end{equation}

i.e. the odds of $i$ continuing the flow in country $j$. 
To ensure the assumptions of our regression procedure are satisfied and to account for the potential correlation of errors within countries or country pairs, we employ Ordinary Least Squares (OLS) with robust clustered standard errors.
Although estimating the parameter $\beta_k$, where $k \in {1, \ldots, 6}$, using standard OLS is common, it may underestimate the actual variance of the parameters \cite{Moulton1990}. This underestimation is due to the potential correlation of errors within the country, such as the correlation between country pairs.
To address this issue, we use OLS with robust clustered standard errors. This involves specifying a clustering variable (in our case $d_{ij}$) that independently identifies each country pair, regardless of the direction. By employing robust clustered standard errors, we can ensure that all assumptions are satisfied and obtain more reliable estimates of the parameters.

After merging the CEPII and IMF datasets, we obtain a network of $172$ nodes and $20468$ edges.
Table~\ref{tab:gm_result} shows the estimated parameters' values.

\begin{table}[!ht]
\centering
\begin{tabular}{rrrrrrr}
  \hline
 & Estimate & Std. Error & t value & Pr($>$$|$t$|$) & CI Lower & CI Upper \\ 
  \hline
(Intercept) & -5.40 & 0.13 & -40.34 & 0.00 & -5.66 & -5.14 \\ 
  $\log(GDP_i)$ & -0.13 & 0.01 & -23.02 & 0.00 & -0.14 & -0.12 \\ 
  $\log(GDP_j)$ & 0.78 & 0.01 & 148.60 & 0.00 & 0.77 & 0.79 \\ 
  $\log(d_{ij})$ & -0.54 & 0.01 & -37.85 & 0.00 & -0.57 & -0.51 \\ 
  $C_{ij}$ & 0.49 & 0.08 & 5.73 & 0.00 & 0.32 & 0.65 \\ 
  $L_{ij}$ & 0.48 & 0.06 & 8.27 & 0.00 & 0.37 & 0.60 \\ 
  $S_{ij}$ & 0.42 & 0.06 & 7.56 & 0.00 & 0.31 & 0.53 \\ 
   \hline
\end{tabular}
\caption{Result of the GM fitting.}
\label{tab:gm_result}
\end{table}

The obtained results indicate that our model explains approximately 59\% of the variance in the dependent variable ($R^2 = 0.59$).

The findings suggest that news flow tends to occur more frequently between countries that are geographically closer to each other. However, the coefficient for distance in the news diffusion model (approximately -0.54) is lower in absolute value compared to what is typically observed in real trade networks (which is approximately 1). This difference can be attributed to the fact that news diffusion does not have the same tangible cost associated with it as in trade. News can travel more easily and quickly across long distances without incurring significant transportation or logistical expenses. Thus, the impact of geographical distance on news diffusion is somewhat attenuated compared to its impact on trade flows.

Additionally, the results indicate that economic power plays a role in news diffusion. Countries with higher GDP are more likely to be the target of the news flow, while they are less likely to be the source of the flow, although this effect is relatively weak ($\beta_1 \approx -0.13$).

Regarding cultural factors, the study reveals that the presence of a shared language has a substantial influence on the transmission of news. This is substantiated by the positive and statistically significant coefficients assigned to the language variables ($L_{ij}$, $S_{ij}$) in the model.

In summary, the analysis indicates that several factors, including geographic proximity, economic power, and cultural aspects like language, play a pivotal role in shaping the dynamics of news diffusion among nations. These findings emphasize the significance of considering these factors when examining the patterns of information flow on an international scale.

\section*{Conclusion}
We adopt a novel approach by utilizing a network-based analysis and a gravity model to investigate the organizational structure and primary drivers of news dissemination in the digital realm. In contrast to prior research, our study incorporates a temporal dimension into the analysis of news diffusion networks, leveraging a vast dataset provided by GDELT.
By analyzing this dataset, we unveil a network that exhibits robust interconnections and reciprocal relationships, highlighting the presence of a core group of nations that hold significant influence in the dissemination of information.

Through the usage of a gravity model, we have uncovered which parameters influence the observed behavior, namely the economic power, the geographic location, and the usage of the same language. 

Possibly limitations of our work regard the model of diffusion that we use. In fact, an edge $(i,j)$ between two countries does not generally represent a direct influence between country $i$ and $j$, i.e. we are not sure that $j$ mention the event because $i$ did.
Moreover, the Multilingual Source-Country dataset provides only an approximation of the geographic country of origin of news outlets.

Although our study uses a different dataset and modeling approach, and despite potential data limitations related to the ongoing Russia-Ukraine conflict, our findings align with prior research on news distribution. These results suggest that the core group of countries that play a prominent role in news distribution remains relatively stable over time. It underscores the continued relevance of economic, geographic, and cultural factors in shaping news dissemination in the digital age.

\section*{Methods}

\subsection*{GDELT Dataset}
 We used the GDELT 2.0 to collect online news articles from January 1 to December 31, 2022, with a 15-minute update resolution. The Global Database of Events, Language, and Tone (GDELT) is an independent, non-profit project providing a vast, updated global news and events database in different languages. The dataset covers news articles written in English and 65 other languages, allowing us to delve deep into non-Western media. The data in GDELT 2.0 is organized into three main tables: Events, Mentions, and the Global Knowledge Graph (GKG). The Events table lists events with related general information. The Mentions table instead contains each news article referring to an event from the Events table. Finally, the GKG table provides detailed information about each article's actors, emotions, and themes. We used the Mentions table to track a story's trajectory and network structure as it flowed through the global media system from country to country. In the Mentions table, each mention (i.e., news article) of an event is given its entry in the dataset. This means an event mentioned in 100 news articles will be listed 100 times in the Mentions table. The dataset resulting from this collection, described in  Table \ref{tab:db_breakdown}, consists of 140 million news articles referencing 51 million unique events. To ensure data accuracy, we excluded news articles from domains not present in the Source-Country dataset, resulting in 125 million news articles and their respective country of origin. We explain GDELT's event identification process in the supplementary information. 
 
 \begin{table}[!ht]
     \centering
     \begin{tabular}{cc}
        \hline
          & Records \\
        \hline
         English Articles & 99.5 M \\
         Non-English Articles & 40.5 M \\
         Total Articles & 140 M \\
         Articles with domains inside Source-Country & 125 M \\
     \end{tabular}
     \caption{Dataset breakdown}
     \label{tab:db_breakdown}
 \end{table}


\subsection*{Multilingual Source-Country Dataset}
This dataset (gdelt-bq.extra.sourcesbycountry) estimates the country of origin for major online news outlets monitored by GDELT, using the primary geographic focus of the outlet over the monitored time frame. GDELT's documentation mentions challenges related to outlets with insufficient coverage volume or geographic emphasis and regional wire services with bureaus in particular countries\cite{gdelt_crossreferencing}. Despite the challenges, the dataset is considered a reasonable approximation of the geographic country of origin for these news outlets and is available on \href{https://blog.gdeltproject.org/multilingual-source-country-crossreferencing-dataset/}{GDELT's website}.

\subsection*{Network construction}

Starting from our collected data, for each event $k$ we consider a directed weighted graph $G_k = (V_k,E_k)$. In particular, $V_k$ is the unique set of countries that mention $k$, while $(i,j) \in E_k$ if country $j$ mentions $k$ successively after country $i$. In more details, consider two successive times $t_i, t_{i+1}$ (we have $t_{i+1}-t_i \geq 15$ min) and let $C_k(t_i),C_k(t_{i+1})$ be the set of countries that mention $k$ at time $t,t+1$ (respectively). If $T_k$ is the lifespan of event $k$, we have that $E_k = \bigcup_{i = 1}^{T_k-1} C_k(t_i)\times C_k(t_{i+1})$. Since it is possible to have multiple edges, we associate to each of them a weight $w_{i,j}^k$ that represents their multiplicity.

This procedure results in a collection of graphs $G_k$, one for each event.
We obtain a final network $G = (V,E)$ defined as

$$
G = \bigcup_{k} G_k
$$

and $w_{ij} = \sum_k w_{ij}^k$, i.e. we overlap all the networks associated with each event. Thus, $G$ is a directed weighted network in which an edge between $(i,j)$ means that $j$ mention an event successively after $i$ a number of times equal to $w_{ij}$. To focus on the relationship among countries, we delete all the loops, i.e. edges that start and end in the same country, obtaining a simple graph.

\subsubsection*{Weighted Clustering Coefficient}
Clustering coefficients are widely used in the network science literature, with the aim of studying the interconnectedness of nodes' neighbours. However, fewer measures have been proposed for weighted directed networks. 

Since one of our aims is to study the organization of news diffusion network, we decided to consider the Clustering Coefficient proposed by Clemente and Grassi \cite{Clemente2018}, which can be defined as follows

\begin{equation}\label{eq:directed_clustering}
    C_i = \frac{0.5 \left[(\bf{W}+\bf{W}^T)(\bf{A}+\bf{A}^T)^2\right]_{ii}}{s_i \left(d_i -1\right) - 2s_i^{\leftrightarrow}}
\end{equation}

where $\bf{A},\bf{W}$ are the unweighted and weighted adjacency matrices of the network, $s_i$ is the strength of node $i$ and $s_i ^{\leftrightarrow}$ is defined as the strength of bilateral arcs, i.e.

$$
s_i ^\leftrightarrow = \sum_{i \ne j} 
a_{ij} a_{ji} \frac{w_{ij} + w_{ji}}{2}.
$$

The numerator of \eqref{eq:directed_clustering} takes into account all directed triangles that node $i$ actually forms with its neighbours, weighted with the average weight of the links connecting a node $i$ to its adjacent $j$ and $k$. The denominator counts the total number of directed triangles that it could form, taking weights properly.

\subsubsection*{Rich-club}
Rich-club refers to a network subgraph made up of the most prominent nodes (from a topological or non-topological point of view) being highly interconnected with each other. Many studies report that these nodes tend to be important for the overall structure and function of the network~\cite{zhou2004rich,van2011rich,ma2015anatomy,cinelli2018rich}. In more detail, given a weighted graph $G = (V,E)$ with binary adjacency matrix $A$ and weighted adjacency matrix $W$ ($W_{i,j} \in \left(0, +\infty \right))$, the topological (i.e. unweighted) measure that allows to detect a rich-club with respect to nodes of degree $k$ is:

\begin{equation}\label{eq:top_rc}
    \phi (k) = \frac{2 m_{>k}}{n_{>k} (n_{>k} -1)},
\end{equation}

where $m_{>k}$ and $n_{>k}$ are the number of links and nodes of the subgraph $G_{>k} \subseteq G$ inducted by nodes with degree higher than $k$.

To consider weighted networks, the presence of a rich-club must be detected with respect to a certain richness parameter $r$ that allows ranking the nodes with respect to it. Common examples of richness parameters can be degree, strength or other measures of centrality~\cite{Opsahl_rich_club,cinelli2019generalized}.

Given such a measure, it is possible to define the weighted rich-club as:

\begin{equation}\label{eq:weighted_rc}
    \phi^w (r) = \frac{W_{>r}}{\sum_{i = 1}^{E_{>r}} W_i^{rank}}
\end{equation}

where $w_i ^{rank} \geq w_{i+1} ^{rank}$, with $i = 1,2,\ldots, |E|$ are the weights of the links ranked in increasing order. 
Thus, \eqref{eq:weighted_rc} measures the fraction of weights shared by the richest nodes compared to the total amount they could share using the strongest links of the network.

However, the two definitions alone, don't assure the existence of a rich-club effect. To measure that, a suitable null model that preserves some properties of the original graph must be considered~\cite{colizza2006detecting}. 
In this context, we are interested in maintaining the strength distribution, therefore we consider a set of randomized networks obtained from $G$ by reshuffling locally the weights of the outgoing links, as proposed by Opsahl et al. \cite{Opsahl_rich_club}. 

We define $\phi_{null} ^{w} (r)$ as the value of \eqref{eq:weighted_rc} computed for the random null model. Thus, we detect a rich-club effect by computing the ratio

\begin{equation}\label{eq:normalized_rc}
    \rho^w (r) = \frac{\phi^w (r)}{\bar{\phi}_{null}^w (r)},
\end{equation}

where $\bar{\phi}_{null}^w (r)$ is the mean across $N$ generated null model. 
A value of $\rho^w (r) > 1$ indicate that rich nodes concentrate their flow towards other rich nodes more than expected in a set (usually containing 100 randomised instances) of null models.
Note that \eqref{eq:normalized_rc} can be applied with no problem to the case of unweighted networks. However, in this case, we consider a null model in which the degree sequence is preferred, using a rewiring algorithm. 

\subsubsection*{HITS Algorithm}
HITS (Hyperlink-Induced Topic Search) \cite{Kleinberg1999} is a web page ranking algorithm used to evaluate their authorities based on their network structure, defined by a directed graph $G = (V,E)$.
The basic idea is to consider a web page to be important if it is linked to other important pages, considering two types of scores associated with each page: {\it Hub} and {\it Authority}. 
In particular a page with a high Hub score points to many good authorities and, vice-versa, a high authority score indicates a page pointed to many good Hubs.

Let's denote with $x(p)$ ($y(p)$) the Authority score (Hub score) of page $p$.
The algorithm initially sets $x(p) = y(p) = 1$ and then updates the values according to the following simple rules:

\begin{enumerate}
    \item For each $p$, $x(p) = \sum_{\{q : (q,p) \in E\}} y(q)$;
    \item For each $p$, $y(p) = \sum_{\{q : (p,q) \in E\}} x(q)$;
    \item Normalize the Authority scores such that $\sum_{p \in V} x(p)^2 = 1$;
    \item Normalize the Hub scores such that $\sum_{p \in V} y(p)^2 = 1$;
\end{enumerate}

Linear Algebra tools assure that the previous algorithm converges to fixed points $\Bar{x}$ and $\Bar{y}$. Moreover, experimental evidence indicates that this convergence is obtained with a limited number of steps.

\subsubsection*{Disparity Index}
The disparity index has been widely used in many fields like economics \cite{hirschman1964paternity},  ecology \cite{simpson1949measurement}, physics \cite{derrida1987statistical} and network science \cite{barthelemy2003spatial, Serrano2009}. It measures the inhomogeneities in the weights at the local level of nodes, and it is defined, for a node $i$ of degree $k$, as

\begin{equation}\label{eq:disparity_index}
    \gamma_i (k) = k \sum_{j} p_{ij} ^ 2.
\end{equation}

Its value ranges between $1$ (perfect homogeneity, i.e. all the links have the same weight) and $k$ (perfect heterogeneity, i.e. only a link carries the whole node strength). 

\bibliography{main}

\begin{thebibliography}{10}
\urlstyle{rm}
\expandafter\ifx\csname url\endcsname\relax
  \def\url#1{\texttt{#1}}\fi
\expandafter\ifx\csname urlprefix\endcsname\relax\def\urlprefix{URL }\fi
\expandafter\ifx\csname doiprefix\endcsname\relax\def\doiprefix{DOI: }\fi
\providecommand{\bibinfo}[2]{#2}
\providecommand{\eprint}[2][]{\url{#2}}

\bibitem{Ahmad2010}
\bibinfo{author}{Ahmad, A.~N.}
\newblock \bibinfo{journal}{\bibinfo{title}{Is twitter a useful tool for
  journalists?}}
\newblock {\emph{\JournalTitle{Journal of Media Practice}}}
  \textbf{\bibinfo{volume}{11}}, \bibinfo{pages}{145--155},
  \doiprefix\url{10.1386/jmpr.11.2.145\_1} (\bibinfo{year}{2010}).
\newblock \eprint{https://doi.org/10.1386/jmpr.11.2.145_1}.

\bibitem{Gentzkow2008}
\bibinfo{author}{Gentzkow, M.} \& \bibinfo{author}{Shapiro, J.~M.}
\newblock \bibinfo{journal}{\bibinfo{title}{Competition and truth in the market
  for news}}.
\newblock {\emph{\JournalTitle{Journal of Economic Perspectives}}}
  \textbf{\bibinfo{volume}{22}}, \bibinfo{pages}{133--154},
  \doiprefix\url{10.1257/jep.22.2.133} (\bibinfo{year}{2008}).

\bibitem{chen2019competition}
\bibinfo{author}{Chen, H.} \& \bibinfo{author}{Suen, W.}
\newblock \bibinfo{journal}{\bibinfo{title}{Competition for attention and news
  quality}}.
\newblock {\emph{\JournalTitle{American Economic Journal: Microeconomics,
  forthcoming}}}  (\bibinfo{year}{2019}).

\bibitem{leetaru2012data}
\bibinfo{author}{Leetaru, K.}
\newblock \emph{\bibinfo{title}{Data mining methods for the content analyst: An
  introduction to the computational analysis of content}}
  (\bibinfo{publisher}{Routledge}, \bibinfo{year}{2012}).

\bibitem{sittar2020dataset}
\bibinfo{author}{Sittar, A.}, \bibinfo{author}{Mladenic, D.} \&
  \bibinfo{author}{Erjavec, T.}
\newblock \bibinfo{title}{A dataset for information spreading over the news}.
\newblock In \emph{\bibinfo{booktitle}{Proceedings of the 23th International
  Multiconference Information Society SiKDD}}, vol. \bibinfo{volume}{100},
  \bibinfo{pages}{5--8} (\bibinfo{year}{2020}).

\bibitem{bodaghi2022theater}
\bibinfo{author}{Bodaghi, A.} \& \bibinfo{author}{Oliveira, J.}
\newblock \bibinfo{journal}{\bibinfo{title}{The theater of fake news spreading,
  who plays which role? a study on real graphs of spreading on twitter}}.
\newblock {\emph{\JournalTitle{Expert Systems with Applications}}}
  \textbf{\bibinfo{volume}{189}}, \bibinfo{pages}{116110}
  (\bibinfo{year}{2022}).

\bibitem{abdullah2011epidemic}
\bibinfo{author}{Abdullah, S.} \& \bibinfo{author}{Wu, X.}
\newblock \bibinfo{title}{An epidemic model for news spreading on twitter}.
\newblock In \emph{\bibinfo{booktitle}{2011 IEEE 23rd international conference
  on tools with artificial intelligence}}, \bibinfo{pages}{163--169}
  (\bibinfo{organization}{IEEE}, \bibinfo{year}{2011}).

\bibitem{schmidt2017anatomy}
\bibinfo{author}{Schmidt, A.~L.} \emph{et~al.}
\newblock \bibinfo{journal}{\bibinfo{title}{Anatomy of news consumption on
  facebook}}.
\newblock {\emph{\JournalTitle{Proceedings of the National Academy of
  Sciences}}} \textbf{\bibinfo{volume}{114}}, \bibinfo{pages}{3035--3039}
  (\bibinfo{year}{2017}).

\bibitem{del2016spreading}
\bibinfo{author}{Del~Vicario, M.} \emph{et~al.}
\newblock \bibinfo{journal}{\bibinfo{title}{The spreading of misinformation
  online}}.
\newblock {\emph{\JournalTitle{Proceedings of the national academy of
  Sciences}}} \textbf{\bibinfo{volume}{113}}, \bibinfo{pages}{554--559}
  (\bibinfo{year}{2016}).

\bibitem{vosoughi2018spread}
\bibinfo{author}{Vosoughi, S.}, \bibinfo{author}{Roy, D.} \&
  \bibinfo{author}{Aral, S.}
\newblock \bibinfo{journal}{\bibinfo{title}{The spread of true and false news
  online}}.
\newblock {\emph{\JournalTitle{science}}} \textbf{\bibinfo{volume}{359}},
  \bibinfo{pages}{1146--1151} (\bibinfo{year}{2018}).

\bibitem{cinelli2021echo}
\bibinfo{author}{Cinelli, M.}, \bibinfo{author}{De~Francisci~Morales, G.},
  \bibinfo{author}{Galeazzi, A.}, \bibinfo{author}{Quattrociocchi, W.} \&
  \bibinfo{author}{Starnini, M.}
\newblock \bibinfo{journal}{\bibinfo{title}{The echo chamber effect on social
  media}}.
\newblock {\emph{\JournalTitle{Proceedings of the National Academy of
  Sciences}}} \textbf{\bibinfo{volume}{118}}, \bibinfo{pages}{e2023301118}
  (\bibinfo{year}{2021}).

\bibitem{spohr2017fake}
\bibinfo{author}{Spohr, D.}
\newblock \bibinfo{journal}{\bibinfo{title}{Fake news and ideological
  polarization: Filter bubbles and selective exposure on social media}}.
\newblock {\emph{\JournalTitle{Business information review}}}
  \textbf{\bibinfo{volume}{34}}, \bibinfo{pages}{150--160}
  (\bibinfo{year}{2017}).

\bibitem{stroud2010polarization}
\bibinfo{author}{Stroud, N.~J.}
\newblock \bibinfo{journal}{\bibinfo{title}{Polarization and partisan selective
  exposure}}.
\newblock {\emph{\JournalTitle{Journal of communication}}}
  \textbf{\bibinfo{volume}{60}}, \bibinfo{pages}{556--576}
  (\bibinfo{year}{2010}).

\bibitem{wanta2004agenda}
\bibinfo{author}{Wanta, W.}, \bibinfo{author}{Golan, G.} \&
  \bibinfo{author}{Lee, C.}
\newblock \bibinfo{journal}{\bibinfo{title}{Agenda setting and international
  news: Media influence on public perceptions of foreign nations}}.
\newblock {\emph{\JournalTitle{Journalism \& mass communication quarterly}}}
  \textbf{\bibinfo{volume}{81}}, \bibinfo{pages}{364--377}
  (\bibinfo{year}{2004}).

\bibitem{kiousis2008international}
\bibinfo{author}{Kiousis, S.} \& \bibinfo{author}{Wu, X.}
\newblock \bibinfo{journal}{\bibinfo{title}{International agenda-building and
  agenda-setting: Exploring the influence of public relations counsel on us
  news media and public perceptions of foreign nations}}.
\newblock {\emph{\JournalTitle{International Communication Gazette}}}
  \textbf{\bibinfo{volume}{70}}, \bibinfo{pages}{58--75}
  (\bibinfo{year}{2008}).

\bibitem{peter2003agenda}
\bibinfo{author}{Peter, J.} \& \bibinfo{author}{De~Vreese, C.~H.}
\newblock \bibinfo{journal}{\bibinfo{title}{Agenda-rich, agenda-poor: A
  cross-national comparative investigation of nominal and thematic public
  agenda diversity}}.
\newblock {\emph{\JournalTitle{International Journal of Public Opinion
  Research}}} \textbf{\bibinfo{volume}{15}}, \bibinfo{pages}{44--64}
  (\bibinfo{year}{2003}).

\bibitem{wallerstein1974wst}
\bibinfo{author}{Wallerstein, I.}
\newblock \emph{\bibinfo{title}{The Modern World-System I: Capitalist
  Agriculture and the Origins of the European World-Economy in the Sixteenth
  Century}} (\bibinfo{publisher}{University of California Press},
  \bibinfo{year}{2011}), \bibinfo{edition}{1} edn.

\bibitem{Chang1998}
\bibinfo{author}{CHANG, T.-K.}
\newblock \bibinfo{journal}{\bibinfo{title}{All countries not created equal to
  be news: World system and international communication}}.
\newblock {\emph{\JournalTitle{Communication Research}}}
  \textbf{\bibinfo{volume}{25}}, \bibinfo{pages}{528--563},
  \doiprefix\url{10.1177/009365098025005004} (\bibinfo{year}{1998}).
\newblock \eprint{https://doi.org/10.1177/009365098025005004}.

\bibitem{kim1996network}
\bibinfo{author}{KIM, K.} \& \bibinfo{author}{BARNETT, G.~A.}
\newblock \bibinfo{journal}{\bibinfo{title}{The determinants of international
  news flow: A network analysis}}.
\newblock {\emph{\JournalTitle{Communication Research}}}
  \textbf{\bibinfo{volume}{23}}, \bibinfo{pages}{323--352},
  \doiprefix\url{10.1177/009365096023003004} (\bibinfo{year}{1996}).
\newblock \eprint{https://doi.org/10.1177/009365096023003004}.

\bibitem{wu2003homogeneity}
\bibinfo{author}{Wu, H.~D.}
\newblock \bibinfo{journal}{\bibinfo{title}{Homogeneity around the world?
  comparing the systemic determinants of international news flow between
  developed and developing countries}}.
\newblock {\emph{\JournalTitle{Gazette (Leiden, Netherlands)}}}
  \textbf{\bibinfo{volume}{65}}, \bibinfo{pages}{9--24} (\bibinfo{year}{2003}).

\bibitem{gdelt_network_2022}
\bibinfo{author}{Guo, L.} \& \bibinfo{author}{Vargo, C.~J.}
\newblock \bibinfo{journal}{\bibinfo{title}{Predictors of international news
  flow: Exploring a networked global media system}}.
\newblock {\emph{\JournalTitle{Journal of Broadcasting \& Electronic Media}}}
  \textbf{\bibinfo{volume}{64}}, \bibinfo{pages}{418--437},
  \doiprefix\url{10.1080/08838151.2020.1796391} (\bibinfo{year}{2020}).
\newblock \eprint{https://doi.org/10.1080/08838151.2020.1796391}.

\bibitem{gravitynews320}
\bibinfo{author}{Grasland, C.}
\newblock \bibinfo{journal}{\bibinfo{title}{International news flow theory
  revisited through a space--time interaction model: Application to a sample of
  320,000 international news stories published through rss flows by 31 daily
  newspapers in 2015}}.
\newblock {\emph{\JournalTitle{International Communication Gazette}}}
  \textbf{\bibinfo{volume}{82}}, \bibinfo{pages}{231--259}
  (\bibinfo{year}{2020}).

\bibitem{Fracasso2014}
\bibinfo{author}{Fracasso, A.}, \bibinfo{author}{Grassano, N.} \&
  \bibinfo{author}{Marzetti, G.~V.}
\newblock \bibinfo{journal}{\bibinfo{title}{The gravity of foreign news
  coverage in the {EU}: Does the euro matter?}}
\newblock {\emph{\JournalTitle{{JCMS}: Journal of Common Market Studies}}}
  \textbf{\bibinfo{volume}{53}}, \bibinfo{pages}{274--291},
  \doiprefix\url{10.1111/jcms.12182} (\bibinfo{year}{2014}).

\bibitem{leetaru2013gdelt}
\bibinfo{author}{Leetaru, K.} \& \bibinfo{author}{Schrodt, P.~A.}
\newblock \bibinfo{title}{Gdelt: Global data on events, location, and tone,
  1979--2012}.
\newblock In \emph{\bibinfo{booktitle}{ISA annual convention}},
  \bibinfo{number}{4}, \bibinfo{pages}{1--49}
  (\bibinfo{organization}{Citeseer}, \bibinfo{year}{2013}).

\bibitem{leetaru2015mining}
\bibinfo{author}{Leetaru, K.~H.}
\newblock \bibinfo{journal}{\bibinfo{title}{Mining libraries: Lessons learned
  from 20 years of massive computing on the world’s information}}.
\newblock {\emph{\JournalTitle{Information Services \& Use}}}
  \textbf{\bibinfo{volume}{35}}, \bibinfo{pages}{31--50}
  (\bibinfo{year}{2015}).

\bibitem{Barrat2004}
\bibinfo{author}{Barrat, A.}, \bibinfo{author}{Barth{\'{e}}lemy, M.},
  \bibinfo{author}{Pastor-Satorras, R.} \& \bibinfo{author}{Vespignani, A.}
\newblock \bibinfo{journal}{\bibinfo{title}{The architecture of complex
  weighted networks}}.
\newblock {\emph{\JournalTitle{Proceedings of the National Academy of
  Sciences}}} \textbf{\bibinfo{volume}{101}}, \bibinfo{pages}{3747--3752},
  \doiprefix\url{10.1073/pnas.0400087101} (\bibinfo{year}{2004}).

\bibitem{Clemente2018}
\bibinfo{author}{Clemente, G.} \& \bibinfo{author}{Grassi, R.}
\newblock \bibinfo{journal}{\bibinfo{title}{Directed clustering in weighted
  networks: A new perspective}}.
\newblock {\emph{\JournalTitle{Chaos, Solitons \& Fractals}}}
  \textbf{\bibinfo{volume}{107}}, \bibinfo{pages}{26--38},
  \doiprefix\url{10.1016/j.chaos.2017.12.007} (\bibinfo{year}{2018}).

\bibitem{xu2010rich}
\bibinfo{author}{Xu, X.-K.}, \bibinfo{author}{Zhang, J.} \&
  \bibinfo{author}{Small, M.}
\newblock \bibinfo{journal}{\bibinfo{title}{Rich-club connectivity dominates
  assortativity and transitivity of complex networks}}.
\newblock {\emph{\JournalTitle{Physical Review E}}}
  \textbf{\bibinfo{volume}{82}}, \bibinfo{pages}{046117}
  (\bibinfo{year}{2010}).

\bibitem{zhou2004rich}
\bibinfo{author}{Zhou, S.} \& \bibinfo{author}{Mondrag{\'o}n, R.~J.}
\newblock \bibinfo{journal}{\bibinfo{title}{The rich-club phenomenon in the
  internet topology}}.
\newblock {\emph{\JournalTitle{IEEE communications letters}}}
  \textbf{\bibinfo{volume}{8}}, \bibinfo{pages}{180--182}
  (\bibinfo{year}{2004}).

\bibitem{Opsahl_rich_club}
\bibinfo{author}{Opsahl, T.}, \bibinfo{author}{Colizza, V.},
  \bibinfo{author}{Panzarasa, P.} \& \bibinfo{author}{Ramasco, J.~J.}
\newblock \bibinfo{journal}{\bibinfo{title}{Prominence and control: The
  weighted rich-club effect}}.
\newblock {\emph{\JournalTitle{Phys. Rev. Lett.}}}
  \textbf{\bibinfo{volume}{101}}, \bibinfo{pages}{168702},
  \doiprefix\url{10.1103/PhysRevLett.101.168702} (\bibinfo{year}{2008}).

\bibitem{cinelli2019generalized}
\bibinfo{author}{Cinelli, M.}
\newblock \bibinfo{journal}{\bibinfo{title}{Generalized rich-club ordering in
  networks}}.
\newblock {\emph{\JournalTitle{Journal of Complex Networks}}}
  \textbf{\bibinfo{volume}{7}}, \bibinfo{pages}{702--719}
  (\bibinfo{year}{2019}).

\bibitem{Kleinberg1999}
\bibinfo{author}{Kleinberg, J.~M.}
\newblock \bibinfo{journal}{\bibinfo{title}{Authoritative sources in a
  hyperlinked environment}}.
\newblock {\emph{\JournalTitle{Journal of the {ACM}}}}
  \textbf{\bibinfo{volume}{46}}, \bibinfo{pages}{604--632},
  \doiprefix\url{10.1145/324133.324140} (\bibinfo{year}{1999}).

\bibitem{Serrano2009}
\bibinfo{author}{Serrano, M.~{\'{A}}.}, \bibinfo{author}{Bogu{\~{n}}{\'{a}},
  M.} \& \bibinfo{author}{Vespignani, A.}
\newblock \bibinfo{journal}{\bibinfo{title}{Extracting the multiscale backbone
  of complex weighted networks}}.
\newblock {\emph{\JournalTitle{Proceedings of the National Academy of
  Sciences}}} \textbf{\bibinfo{volume}{106}}, \bibinfo{pages}{6483--6488},
  \doiprefix\url{10.1073/pnas.0808904106} (\bibinfo{year}{2009}).

\bibitem{conte2021cepii}
\bibinfo{author}{Conte, M.}, \bibinfo{author}{Cotterlaz, P.},
  \bibinfo{author}{Mayer, T.} \emph{et~al.}
\newblock \bibinfo{journal}{\bibinfo{title}{The cepii gravity database}}.
\newblock {\emph{\JournalTitle{CEPII: Paris, France}}}  (\bibinfo{year}{2021}).

\bibitem{IMF}
\bibinfo{title}{International monetary fund}.
\newblock
  \bibinfo{howpublished}{\url{https://www.imf.org/external/datamapper/NGDPD@WEO/OEMDC/ADVEC/WEOWORLD}}.

\bibitem{Moulton1990}
\bibinfo{author}{Moulton, B.~R.}
\newblock \bibinfo{journal}{\bibinfo{title}{An illustration of a pitfall in
  estimating the effects of aggregate variables on micro units}}.
\newblock {\emph{\JournalTitle{The Review of Economics and Statistics}}}
  \textbf{\bibinfo{volume}{72}}, \bibinfo{pages}{334},
  \doiprefix\url{10.2307/2109724} (\bibinfo{year}{1990}).

\bibitem{gdelt_crossreferencing}
\bibinfo{author}{GDELT}.
\newblock \bibinfo{title}{Announcing new source country cross-referencing
  dataset}.
\newblock
  \bibinfo{howpublished}{\url{https://blog.gdeltproject.org/announcing-new-source-country-crossreferencing-dataset/}}
  (\bibinfo{year}{2014}).

\bibitem{van2011rich}
\bibinfo{author}{Van Den~Heuvel, M.~P.} \& \bibinfo{author}{Sporns, O.}
\newblock \bibinfo{journal}{\bibinfo{title}{Rich-club organization of the human
  connectome}}.
\newblock {\emph{\JournalTitle{Journal of Neuroscience}}}
  \textbf{\bibinfo{volume}{31}}, \bibinfo{pages}{15775--15786}
  (\bibinfo{year}{2011}).

\bibitem{ma2015anatomy}
\bibinfo{author}{Ma, A.}, \bibinfo{author}{Mondrag{\'o}n, R.~J.} \&
  \bibinfo{author}{Latora, V.}
\newblock \bibinfo{journal}{\bibinfo{title}{Anatomy of funded research in
  science}}.
\newblock {\emph{\JournalTitle{Proceedings of the National Academy of
  Sciences}}} \textbf{\bibinfo{volume}{112}}, \bibinfo{pages}{14760--14765}
  (\bibinfo{year}{2015}).

\bibitem{cinelli2018rich}
\bibinfo{author}{Cinelli, M.}, \bibinfo{author}{Ferraro, G.} \&
  \bibinfo{author}{Iovanella, A.}
\newblock \bibinfo{journal}{\bibinfo{title}{Rich-club ordering and the dyadic
  effect: Two interrelated phenomena}}.
\newblock {\emph{\JournalTitle{Physica A: Statistical Mechanics and its
  Applications}}} \textbf{\bibinfo{volume}{490}}, \bibinfo{pages}{808--818}
  (\bibinfo{year}{2018}).

\bibitem{colizza2006detecting}
\bibinfo{author}{Colizza, V.}, \bibinfo{author}{Flammini, A.},
  \bibinfo{author}{Serrano, M.~A.} \& \bibinfo{author}{Vespignani, A.}
\newblock \bibinfo{journal}{\bibinfo{title}{Detecting rich-club ordering in
  complex networks}}.
\newblock {\emph{\JournalTitle{Nature physics}}} \textbf{\bibinfo{volume}{2}},
  \bibinfo{pages}{110--115} (\bibinfo{year}{2006}).

\bibitem{hirschman1964paternity}
\bibinfo{author}{Hirschman, A.~O.}
\newblock \bibinfo{journal}{\bibinfo{title}{The paternity of an index}}.
\newblock {\emph{\JournalTitle{The American economic review}}}
  \textbf{\bibinfo{volume}{54}}, \bibinfo{pages}{761--762}
  (\bibinfo{year}{1964}).

\bibitem{simpson1949measurement}
\bibinfo{author}{Simpson, E.~H.}
\newblock \bibinfo{journal}{\bibinfo{title}{Measurement of diversity}}.
\newblock {\emph{\JournalTitle{nature}}} \textbf{\bibinfo{volume}{163}},
  \bibinfo{pages}{688--688} (\bibinfo{year}{1949}).

\bibitem{derrida1987statistical}
\bibinfo{author}{Derrida, B.} \& \bibinfo{author}{Flyvbjerg, H.}
\newblock \bibinfo{journal}{\bibinfo{title}{Statistical properties of randomly
  broken objects and of multivalley structures in disordered systems}}.
\newblock {\emph{\JournalTitle{Journal of Physics A: Mathematical and
  General}}} \textbf{\bibinfo{volume}{20}}, \bibinfo{pages}{5273}
  (\bibinfo{year}{1987}).

\bibitem{barthelemy2003spatial}
\bibinfo{author}{Barthelemy, M.}, \bibinfo{author}{Gondran, B.} \&
  \bibinfo{author}{Guichard, E.}
\newblock \bibinfo{journal}{\bibinfo{title}{Spatial structure of the internet
  traffic}}.
\newblock {\emph{\JournalTitle{Physica A: statistical mechanics and its
  applications}}} \textbf{\bibinfo{volume}{319}}, \bibinfo{pages}{633--642}
  (\bibinfo{year}{2003}).

\end{thebibliography}

\section*{Acknowledgements}

The work is supported by IRIS Infodemic Coalition (UK government, grant no. SCH-00001-3391), 
SERICS (PE00000014) under the NRRP MUR program funded by the European Union - NextGenerationEU, project CRESP from the Italian Ministry of Health under the program CCM 2022, and PON project “Ricerca e Innovazione” 2014-2020.

\section*{Author contributions statement}

N.D.M. and W.Q. conceived the experiments;
S.A., N.D.M. conducted the experiments;
S.A., N.D.M., M.C. and W.Q. analysed the results;
All authors reviewed the manuscript. 

\section*{Supplementary Information}

\subsection*{GDELT event identification process}
GDELT identifies events by monitoring news reports and using natural language processing (NLP) techniques to extract quotes, people, organizations, and locations mentioned in the text.
The number of unique events in GDELT's dataset, 51 million, may appear unusual at first glance. To clarify this, we provide an example based on GDELT's documentation. Suppose we have a short text: "The United States criticized Russia yesterday for deploying its troops in Crimea, in which a recent clash with its soldiers left ten civilians injured." In this case, GDELT would generate three distinct events: "US CRITICIZES RUSSIA," "RUSSIA TROOP-DEPLOY UKRAINE (CRIMEA)," and "RUSSIA MATERIAL-CONFLICT CIVILIANS (CRIMEA)." This method allows GDELT to capture multiple events within a single news article, leading to a higher number of unique events in our dataset.

\subsection*{Interactions between mutual edges}
Consider the directed weighted network of news diffusion $G$. To understand if mutual edges share the same amount of information, we normalize the weights of mutual edges of $G$ using

\begin{equation}\label{eq:transition_prob}
    p_{ij} = \frac{w_{ij}}{\sum_{k = 1}^{deg(i)} w_{ik}},
\end{equation}

i.e. we compute the transition probabilities between countries.
Therefore, if $w_{ij} \approx w_{ji}$ there is evidence that the flow of information is similar between source and target (and vice-versa).

\begin{figure}[!ht]
    \centering
    \includegraphics[scale = 0.8]{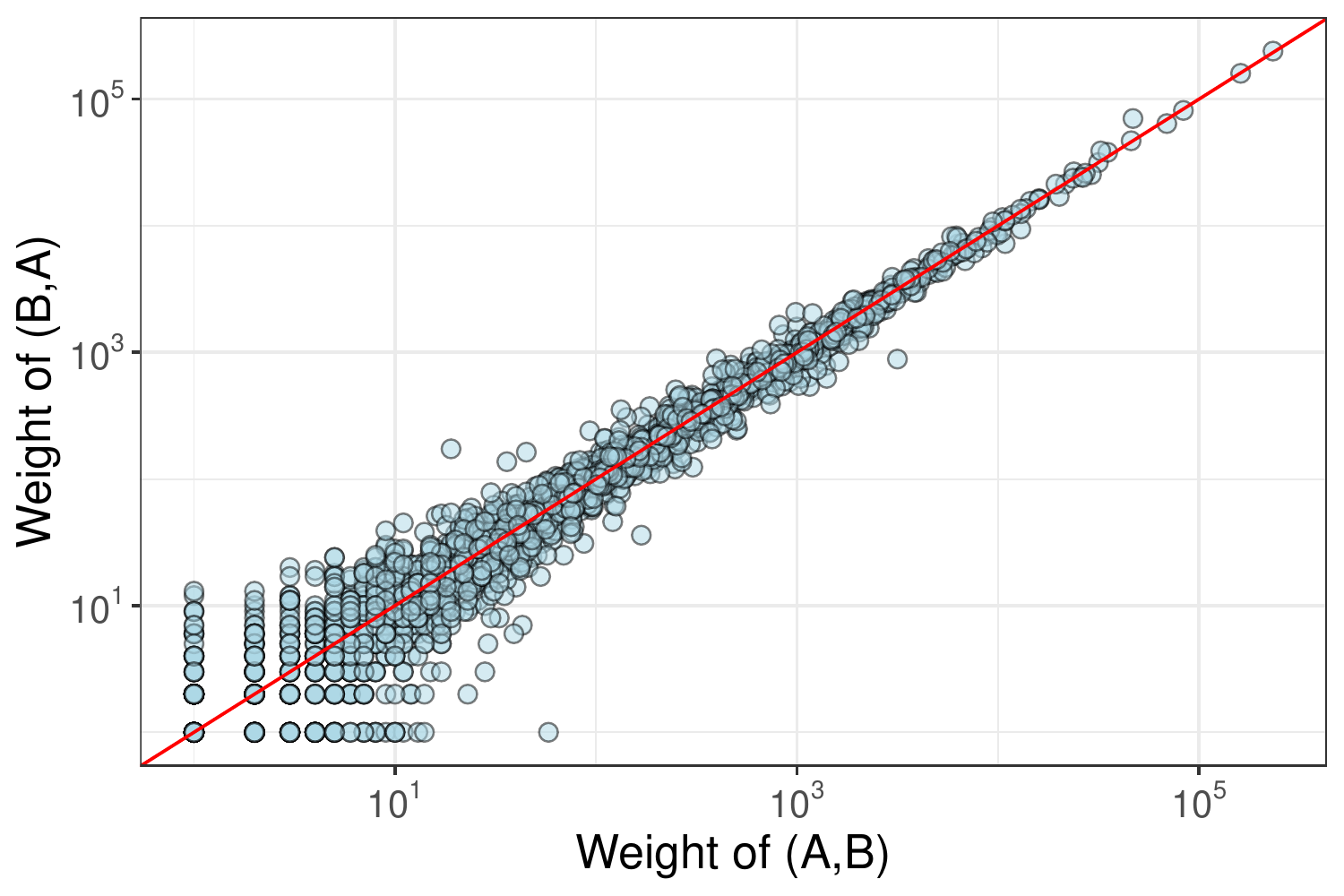}
    \caption{Comparison between weights of mutual edges. Only a sample of $2000$ points is considered for better graphics reasons.}
    \label{fig:mutual_edges}
\end{figure}

Figure~\ref{fig:mutual_edges} shows that in general, we can  deduce that connected countries do not differ much in how they share information, especially for high weights links.

\subsubsection*{Core of countries}

To try to understand how countries are organised in the network, we consider a notion of distance between nodes using the idea that "close" countries are the ones that have more interactions (i.e. a high weight). Therefore, we update the weights of $G$ considering:

\begin{equation}\label{eq:inverse_weights}
    d_{ij} = \frac{1}{w_{ij}}
\end{equation}

obviously, this does not correspond to a distance in a mathematical sense (for example, $d_{ij} \ne d_{ji}$ in general), but this transformation allows us to compute distances taking into account the closeness of countries in terms of news diffusion.

\begin{figure}[!ht]
    \centering
    \includegraphics[width = \linewidth]{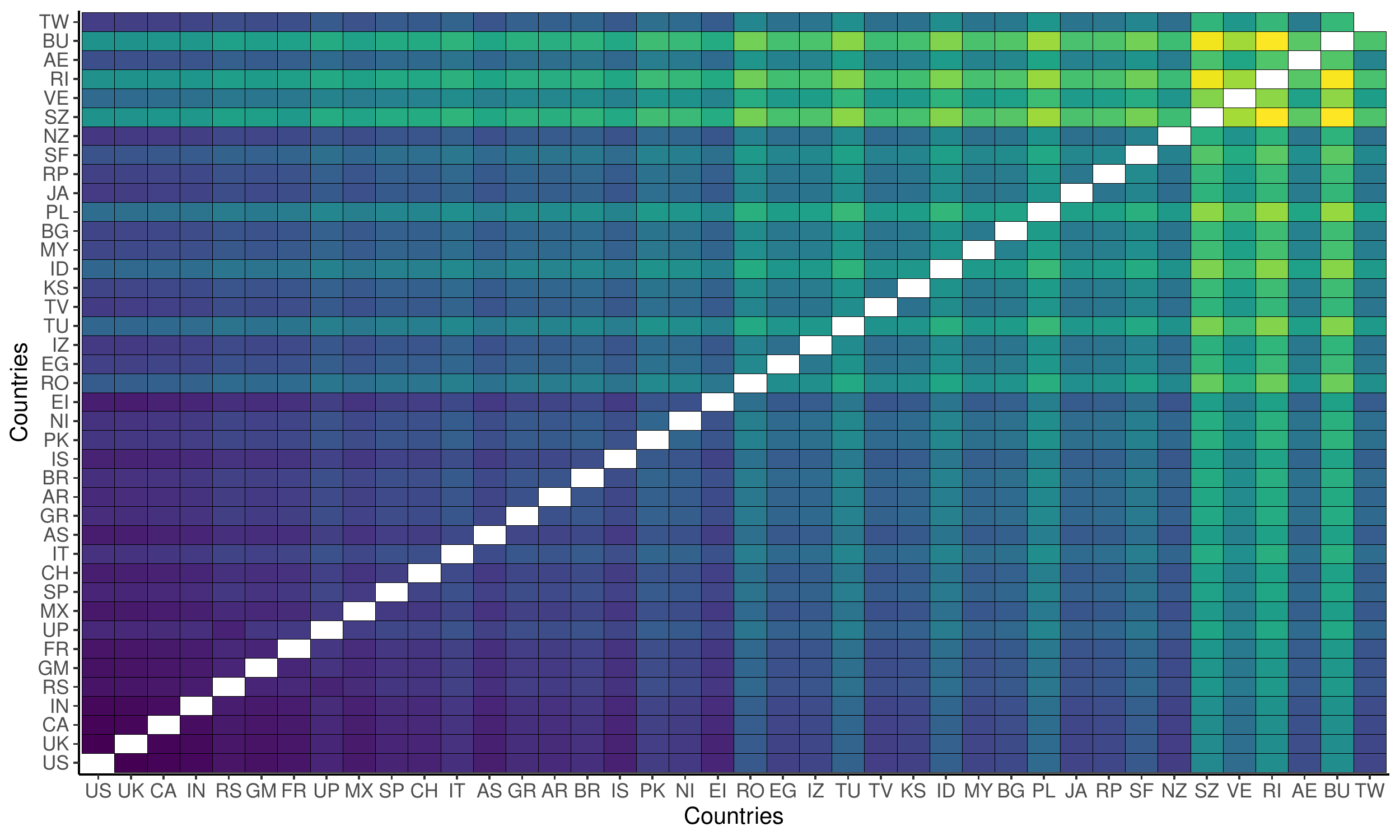}
    \caption{Heatmap of shortest paths among 40 countries with the highest out-strength. We use \eqref{eq:inverse_weights} as weights. The countries are ordered using out-strength.}
    \label{fig:heatmap}
\end{figure}

Figure~\ref{fig:heatmap} shows a heatmap of the shortest distances computed with the weights $\bf{d}$. The heatmap order followed the out-strength value.

Interestingly, countries with high out-strength tend to have low distances in the network. This suggests the existence of a core of countries in which information flows preferentially.

\end{document}